\documentclass{emulateapj}
\usepackage{amsmath}

\newcommand\bB{\boldsymbol{B}}
\newcommand\bb{\boldsymbol{b}}
\newcommand\bk{\boldsymbol{k}}
\newcommand\br{\boldsymbol{r}}
\newcommand\bu{\boldsymbol{u}}
\newcommand\bxi{\boldsymbol{\xi}}

\begin{document} 

\title{Rayleigh-Taylor instabilities with sheared magnetic fields}

\author{M. S. Ruderman}
\affil{Solar Physics and Space Plasma Research Centre (SP$^2$RC), University of
Sheffield, Hicks Building, Hounsfield Road, Sheffield S3 7RH, UK and Space
Research Institute (IKI) Russian Academy of Sciences, Moscow, Russia}
\author{J. Terradas, J. L. Ballester}
\affil{Departament de
F\'{\i}sica, Universitat de les Illes Balears, 07122 Palma de Mallorca, Spain}


\begin{abstract}Magnetic Rayleigh-Taylor (MRT) instabilities may play a relevant role in many astrophysical problems. In this work the effect of magnetic shear on the growth rate of the MRT instability is investigated. The eigenmodes of an
interface and a slab model under the presence of gravity are analytically
calculated  assuming that the orientation of the magnetic field changes in the
equilibrium, i.e., there is magnetic shear. We solve the linearised
magnetohydrodynamic (MHD) equations in the incompressible regime.We find that
the growth rate is bounded under the presence of magnetic shear. We have derived
simple analytical expressions for the maximum growth rate, corresponding to the
most unstable mode of the system. These expressions provide the explicit
dependence of the growth rate on the various equilibrium parameters. For small angles the growth time is linearly proportional to the shear angle, and in this regime the single interface problem and the slab problem tend to the same
result. On the contrary, in the limit of large angles and for the interface
problem the growth time is essentially independent of the shear angle. In this regime we have also been able to calculate an approximate expression for the growth time for the slab configuration. Magnetic shear can have a strong effect on the growth rates of the instability. As an application of the results found in this paper we have indirectly determined the shear angle in solar prominence threads using their lifetimes and the estimation of the Alfv\'en speed of the structure.
\end{abstract} 



\maketitle

\section{Introduction}
\label{sec:intro}

The magnetic Rayleigh-Taylor (MRT) instability is important in many
astrophysical systems. Some examples are buoyant magnetised bubbles identified
in clusters of galaxies, see \citet[][]{robinson2004,jones2005} for studies in
2D, and \citet[][]{oneil2009} for 3D configurations. MRT instabilities also
manifest themselves in shells of young supernova remnants, this has been
investigated by  \citet[][]{jun1995} in 2 and 3D Cartesian configurations and by
\citet[][]{jun1996} in 3D using spherical coordinates. \citet{bucc2004} have
numerically investigated the development of the MRT instability at the interface
between an expanding pulsar wind nebula and its surrounding supernova remnant.
\citet{stonegardiner2007} studied the behaviour of magnetic Rayleigh-Taylor
instability in three dimensions with special focus on the structure and dynamics
of the nonlinear evolution of the system. They analysed various configurations
including the situation in which magnetic fields  change direction at the
interface between the two fluids.  \citet{stonegardiner2007} used the MRT
instability to explain the structure of the optical filaments observed in the
Crab nebula. 

In laboratory plasmas the possible stabilising effect by a force-free magnetic field has been studied in the past by many authors \citep[see for
example][]{goedbloed1971a,goedbloed1971b,goedbloed1971c,GOEPOE2004} using the
single interface problem and the slab problem and applying vacuum conditions at
some of the boundaries. \citet{yang2011} have studied the magnetic field
transition layer effects on the MRT instability with continuous magnetic field
and density profiles and have found that the linear growth rate of the MRT
instability increases with the thickness of the magnetic field transition layer,
especially for the case of small thickness. Recently, \citet{zhangetal2012} have used the ideal MHD model to study the effect of magnetic shear in a finite slab representing a magnetic liner, which is a device used in experiments with fusion plasmas. These authors have found that magnetic shear reduces the MRT growth rate in general. 

The emergence of magnetic flux from the solar interior and the formation of flux
tubes is another example where MRT instabilities are relevant. For example,
\citet[][]{isobe2005,isobe2006} proposed that the MRT instability is a possible
cause of the filamentary structure in mass and current density in the emerging
flux regions. In the solar atmosphere, \citet{ryuetal10} suggested that several
dynamic processes taking place in prominences are most probably related to
magnetic Rayleigh-Taylor instabilities. Along this line of work,
\citet{hillier2011,hillier2012a,hillier2012b} have performed three-dimensional
magnetohydrodynamic simulations to investigate the nonlinear evolution of the
Kippenhahn-Shl\"uter prominence model to the MRT instability. 

The fine structure of solar prominences reveals the presence of magnetic
threads. These structures are quite thin,  of the order of $100\,\rm km$,
aligned with the magnetic field and, in many cases, they seem to lie horizontally
with respect to the photosphere \citep[see][for recent results
about the formation of these structures]{devore2012,devore2013}.
\citet{terretal2012} have considered the possible link between magnetic
Rayleigh-Taylor instabilities and the short thread lifetimes. In that work a
slab model permeated by a horizontal magnetic field was
considered. The growth rates of the unstable modes and the thresholds for
stability were determined analytically. In the present paper we extend the study
to the situation with a sheared magnetic field in which the magnetic field
changes its direction at the interfaces of the plasma slab.
To understand the results in the slab model we describe first the
effect of shear at a single plasma interface. Magnetic shear
introduces changes in the growth rates of the unstable modes that might be
relevant regarding the lifetime of threads. In this work we analytically
calculate these growth rates and perform a detailed analysis of their dependence on the equilibrium parameters.

\section{Problem formulation}
\label{sec:formul}

To describe the plasma motion we use the linearised ideal MHD equation for
incompressible plasmas

\begin{equation}
\nabla\cdot\bxi = 0,
\label{eq:2.1}
\end{equation}
\begin{equation}
\rho\frac{\partial^2\bxi}{\partial t^2} = -\nabla p + 
   \frac1{\mu_0}(\nabla\times\bb)\times\bB,
\label{eq:2.2}
\end{equation}
\begin{equation}
\bb = \nabla\times(\bxi\times\bB) .
\label{eq:2.3}
\end{equation}
Here $\bxi$ is the plasma displacement related to the plasma velocity
$\bu$ by $\bu = \partial\bxi/\partial t$\/, $p$ the pressure
perturbation, and $\bb$ the magnetic field perturbation; $\bB$ is the
background magnetic field, $\rho$ the plasma density assumed to be piecewise
constant, and $\mu_0$ the magnetic permeability of free space. When deriving Eqs.~(\ref{eq:2.1})--(\ref{eq:2.3}) we have assumed that the equilibrium is static and current-free, e.g.\ $\nabla\times\bB = 0$.

\begin{figure} \centerline{\includegraphics[width=17pc]{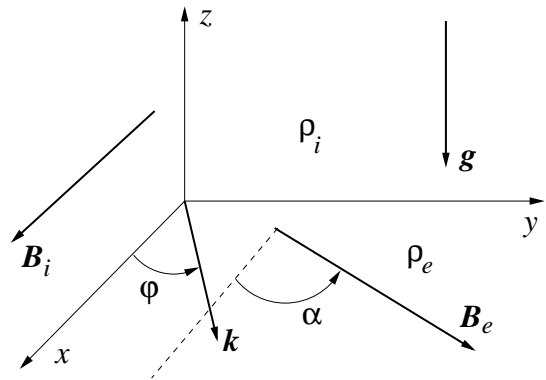}}
\caption{Sketch of single magnetic interface.} \label{inter} 
\end{figure}

In what follows we consider two equilibrium states. In the first one there are
two semi-infinite regions separated by the $xy$\/-plane in Cartesian coordinates
$x$\/, $y$\/, $z$ with the $z$\/-axis in the vertical direction, see Fig.~\ref{inter}. The plasma density and background magnetic field are constant in the two regions and they are given by
\begin{equation}
\rho = \left\{\begin{array}{ll} \rho_e, & z < 0, \\ \rho_i, & z > 0, \end{array}\right. \quad 
\bB = \left\{\begin{array}{ll} \bB_e, & z < 0, \\ \bB_i, & z > 0. \end{array}\right.
\label{eq:2.4}
\end{equation}
The background magnetic field is assumed to be parallel to the $xy$\/-plane. The equilibrium pressure $P$ is defined by the equation  
\begin{equation}
\frac{dP}{dz} = -g\rho ,
\label{eq:2.5}
\end{equation}
where $g$ is the gravity acceleration. The total pressure, magnetic plus
kinetic, has to be continuous at $z = 0$. The solution to Eq.~(\ref{eq:2.5}) satisfying this condition is 
\begin{equation}
P = \left\{\begin{array}{ll} \displaystyle P_0 - \frac{B_e^2}{2\mu_0} 
   - g\rho_e z, & z < 0, \vspace*{2mm}\\ 
\displaystyle  P_0 - \frac{B_i^2}{2\mu_0}  
   - g\rho_i z, & z > 0 , \end{array}\right.
\label{eq:2.6}
\end{equation}
where $P_0$ is an arbitrary constant. We call this equilibrium state the
single magnetic interface.

\begin{figure} \centerline{\includegraphics[width=17pc]{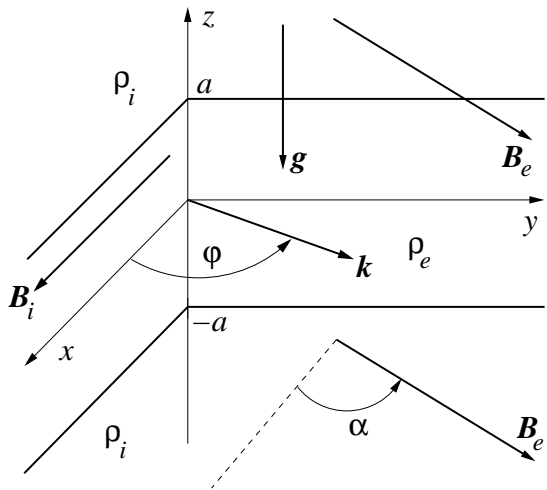}}
\caption{Sketch of magnetic slab.} \label{slab} 
\end{figure}

In the second equilibrium state there are three regions separated by horizontal
planes at $z = \pm a$\/, see Fig.~\ref{slab}. The plasma density and background magnetic field are the same in two semi-infinite regions and they are given by
\begin{equation}
\rho = \left\{\begin{array}{ll} \rho_e, & z < -a, \\ \rho_i, & |z| < a, \\
   \rho_e, & z > a, \end{array}\right. \quad 
\bB = \left\{\begin{array}{ll} \bB_e, & z < -a, \\ 
   \bB_i, & |z| < a, \\ \bB_e, & z > a. \end{array}\right.
\label{eq:2.7}
\end{equation}
The background magnetic field is once again assumed to be parallel to the
$xy$\/-plane. The total pressure has to be continuous at $z = \pm a$\/. The
solution to Eq.~(\ref{eq:2.5}) satisfying this condition is 
\begin{equation}
P = \left\{\begin{array}{ll} \displaystyle P_0 - \frac{B_e^2}{2\mu_0}\, 
   +\, ga(\rho_i - \rho_e) - g\rho_e z, & z < -a, \vspace*{2mm}\\ 
\displaystyle P_0 - \frac{B_i^2}{2\mu_0} - g\rho_i z, & |z| < a , \vspace*{2mm}\\
\displaystyle P_0 - \frac{B_e^2}{2\mu_0}\, -\, 
   ga(\rho_i - \rho_e) - g\rho_e z, & z > a. \end{array}\right.
\label{eq:2.8}
\end{equation}
This second configuration is called the magnetic slab.

At the boundaries separating regions with different plasma densities and
background magnetic field the plasma displacement in the $z$\/-direction and the
Lagrangian perturbation of the total pressure have to be continuous. Hence, we
have two boundary conditions,  
\begin{equation}
[\,\xi_z\,] = 0, \quad [\,p_T - g\rho\xi_z\,] = 0, 
\label{eq:2.9}
\end{equation}
where the square brackets denote the jump of a quantity across a discontinuity,
and $p_T = p + \bB\cdot\bb/\mu_0$ is the perturbation of the total
pressure. When deriving the second boundary condition we have used
Eq.~(\ref{eq:2.5}). The boundary conditions (\ref{eq:2.9}) have to be satisfied
at $z = 0$ in the case of the single magnetic interface, and at $z = \pm a$ in the case of the magnetic slab. One additional boundary condition is that all
perturbations have to vanish as $|z| \to \infty$\/. 

Equations~(\ref{eq:2.1})--(\ref{eq:2.3}) together with the boundary
conditions~(\ref{eq:2.9}) are used in the next section to derive the dispersion
relations determining the stability of the two equilibrium configurations. 

\section{Derivation of the dispersion relations}
\label{sec:dispers}

We Fourier-analyse the perturbations of all quantities and take them proportional to $\exp[i(\bk\cdot\br - \omega t)]$, where $\bk = (k_x,k_y,0)$ and $\br = (x,y,z)$. Then Eqs.~(\ref{eq:2.1})--(\ref{eq:2.3}) reduce to
\begin{equation}
\frac{d\xi_z}{dz} + i\bk\cdot\bxi = 0,
\label{eq:3.1}
\end{equation}
\begin{equation}
\rho\omega^2\bxi_\perp = i\bk p_T - \frac i{\mu_0}\bb_\perp(\bk\cdot\bB),
\label{eq:3.2}
\end{equation}
\begin{equation}
\rho\omega^2\xi_z = \frac{dp_T}{dz} - \frac i{\mu_0}b_z(\bk\cdot\bB),
\label{eq:3.3}
\end{equation}
\begin{equation}
\bb = i(\bk\cdot\bB)\bxi , 
\label{eq:3.4}
\end{equation}
where $\bxi_\perp$ and $\bb_\perp$ are the components of the plasma
displacement and magnetic field perturbation orthogonal to the $z$\/-axis.
Eliminating all the variables from Eqs.~(\ref{eq:3.1})--(\ref{eq:3.4}) in favour of $\xi_z$ we obtain the equation for this variable,
\begin{equation}
\frac{d^2\xi_z}{dz^2} - k^2\xi_z = 0.
\label{eq:3.5}
\end{equation}
In addition, we obtain the expression of $p_T$ in terms of $\xi_z$\/, 
\begin{equation}
p_T = \frac{\rho(\omega^2 - \omega_A^2)}{k^2}\frac{d\xi_z}{dz},
\label{eq:3.6}
\end{equation}
where $\omega_A$ is the Alfv\'en frequency defined by
\begin{equation}
\omega_A^2 = \frac{(\bk\cdot\bB)^2}{\mu_0\rho}.
\label{eq:3.6a}
\end{equation}
This expression enables us to rewrite the boundary conditions~(\ref{eq:2.9}) in
terms of $\xi_z$\/ as
\begin{equation}
[\,\xi_z\,] = 0, \quad \left[\rho(\omega^2 - \omega_A^2)\frac{d\xi_z}{dz} - 
   g\rho k^2\xi_z\right] = 0. 
\label{eq:3.7}
\end{equation}

\subsection{Dispersion relation for a single magnetic interface}
\label{subsec:dispers_single}

In the case of a single magnetic interface the solution to Eq.~(\ref{eq:3.5}),
satisfying the first boundary condition in Eq.~(\ref{eq:3.7}) at $z = 0$
and decaying as $|z| \to \infty$\/, is given, with the accuracy up to an
arbitrary multiplicative constant, by
\begin{equation}
\xi_z = \left\{\begin{array}{ll} e^{kz}, & z < 0, \vspace*{1mm}\\ 
   e^{-kz}, & z > 0. \end{array}\right.
\label{eq:3.8}
\end{equation}
Substituting this solution in the second boundary condition in
Eq.~(\ref{eq:3.7}) we obtain the following dispersion relation
\begin{equation}
\omega^2 = \frac{\rho_e\omega_{Ae}^2 + \rho_i\omega_{Ai}^2 + gk(\rho_e - \rho_i)}
   {\rho_e + \rho_i} .
\label{eq:3.9}
\end{equation}
This is the well-known dispersion equation for the interface problem in an
incompressible fluid (\citealp{Chandrasekhar1961}). When $g = 0$ this is the
dispersion equation for surface waves on a magnetic interface (e.g.\
\citealp{Roberts1981a}). On the other hand, when there is no magnetic field,
this dispersion equation determines the Rayleigh-Taylor instability of the
interface between two incompressible fluids (\citealp{Rayleigh1883,Taylor1950}).

\subsection{Dispersion relation for the magnetic slab}
\label{subsec:dispers_slab}

Now we proceed to the derivation of the dispersion equation for the magnetic slab. The general solution to Eq.~(\ref{eq:3.5}) continuous at $z = \pm a$ and
decaying as $|z| \to \infty$ is
\begin{equation}
\xi_z = \left\{\begin{array}{ll} \left[C_1\cosh(ka) - C_2\sinh(ka)\right]e^{k(z+a)}, & z < -a, \vspace*{1.5mm}\\
\hphantom{[} C_1\cosh(kz) + C_2\sinh(kz), & |z| < a, \vspace*{1.5mm}\\  
\left[C_1\cosh(ka) + C_2\sinh(ka)\right]e^{-k(z-a)}, & z > a, 
\end{array}\right.
\label{eq:3.10}
\end{equation}
where $C_1$ and $C_2$ are arbitrary constants. Substituting this solution in the
second boundary condition in Eq.~(\ref{eq:3.7}) we obtain two equations,
\begin{equation}
A_{11}C_1 - A_{12}C_2 = 0, \quad A_{21}C_1 + A_{22}C_2 = 0,
\label{eq:3.11}
\end{equation}
where
\begin{equation}
\begin{array}{l}
A_{11} = \rho_e(\omega^2 - \omega_{Ae}^2) + gk(\rho_i - \rho_e)   
   \vspace*{1.5mm}\\ 
\hspace*{32mm} +\,\rho_i(\omega^2 - \omega_{Ai}^2)\tanh(ka),\vspace*{1.5mm}\\
A_{12} = \big[\rho_e(\omega^2 - \omega_{Ae}^2) + 
   gk(\rho_i - \rho_e)\big]\tanh(ka) \vspace*{1.5mm}\\
\hspace*{46mm} +\,\rho_i(\omega^2 - \omega_{Ai}^2), \vspace*{1.5mm}\\ 
A_{21} = \rho_e(\omega^2 - \omega_{Ae}^2) - gk(\rho_i - \rho_e)
   \vspace*{1.5mm}\\ 
\hspace*{32mm} +\,\rho_i(\omega^2 - \omega_{Ai}^2)\tanh(ka),\vspace*{1.5mm}\\
A_{22} = \big[\rho_e(\omega^2 - \omega_{Ae}^2) - 
   gk(\rho_i - \rho_e)\big]\tanh(ka) \vspace*{1.5mm}\\
\hspace*{46mm} +\,\rho_i(\omega^2 - \omega_{Ai}^2).
\end{array} 
\label{eq:3.12}
\end{equation}
The system~(\ref{eq:3.11}) of linear homogeneous equations for $C_1$ and $C_2$ 
has non-trivial solutions when its determinant is zero. This condition is
written as $A_{11}A_{22} + A_{12}A_{21}= 0$. After some algebra this equation
gives
\begin{eqnarray}
&&\omega^4\big[2\rho_e\rho_i + \big(\rho_e^2 + \rho_i^2\big)\tanh(2ka)\big] - 
   2\omega^2\big[\rho_e\rho_i\big(\omega_{Ae}^2 \hphantom{xxx}\nonumber\\
&& \hphantom{xx} +\,\omega_{Ai}^2\big) + \big(\rho_e^2\omega_{Ae}^2 +
    \rho_i^2\omega_{Ai}^2\big)\tanh(2ak)\big] \nonumber\\
&& \hphantom{xx} +\, 2\rho_e\rho_i\omega_{Ae}^2\omega_{Ai}^2 +
   \big(\rho_e^2\omega_{Ae}^4 + \rho_i^2\omega_{Ai}^4\big)\tanh(2ak) \nonumber\\
&& \hphantom{xx} -\, g^2 k^2(\rho_e - \rho_i)^2\tanh(2ak) = 0. 
\label{eq:3.13}
\end{eqnarray}
The two solutions to this dispersion equation are $\omega_+^2$ and $\omega_-^2$ given by 
\begin{equation}
\omega_\pm^2 = \frac{F \pm G}H ,
\label{eq:3.14}
\end{equation}
where
\begin{eqnarray}
F &=& \rho_i\rho_e\big(\omega_{Ai}^2 + \omega_{Ae}^2\big)\cosh(2ka)\nonumber\\
&+& \big(\rho_i^2\omega_{Ai}^2 + \rho_e^2\omega_{Ae}^2\big)\sinh(2ka) ,
\label{eq:3.15}
\end{eqnarray}
\begin{eqnarray}
G &=& \Big\{\rho_i^2\rho_e^2(\omega_{Ai}^2 - \omega_{Ae}^2)^2 +
   g^2 k^2(\rho_i - \rho_e)^2  \nonumber\\
&\times& \big[\big(\rho_i^2 + \rho_e^2\big)\sinh^2(2ka) +
   \rho_i\rho_e\sinh(4ka)\big]\Big\}^{1/2} , \hphantom{xx}
\label{eq:3.16}
\end{eqnarray}
\begin{equation}
H = 2\rho_i\rho_e\cosh(2ka) + \big(\rho_i^2 + \rho_e^2\big)\sinh(2ka) .
\label{eq:3.17}
\end{equation}
When $g = 0$ the dispersion relation given by Eq.~(\ref{eq:3.14}) describes
waves in a magnetic slab (e.g.\ \citealp{Parker1974,Edwin1982}). The plus sign corresponds to kink waves where $\xi_z(z)$ is an
even function of $z$\/, while the minus sign corresponds to sausage waves where
$\xi_z(z)$ is an odd function of $z$\/. Although, when $g \neq 0$, $\xi_z(z)$ is
neither odd nor even in both perturbation modes described by
Eq.~(\ref{eq:3.14}), we will still use the name ``kink'' for modes with the plus
sign, and ``sausage'' for modes with the minus sign. It can be also shown that
Eq.~(\ref{eq:3.14}) in the absence of magnetic shear reduces to Eq.~(22) in
\citet{terretal2012}. 

\section{Investigation of stability}
\label{sec:stability}

Here we use the dispersion equations derived in the previous section to study
the stability of a single magnetic interface and a magnetic slab. 

\subsection{Stability of a single magnetic interface}
\label{subsec:stab_single}

Without loss of generality we can choose the $x$\/-axis in the direction of the
vector $\bB_i$\/. It is convenient to introduce the angle $\varphi$ between
the $x$\/-axis and the wave vector $\bk$\/. Then we write $\bk =
k(\cos\varphi,\sin\varphi,0)$, and we also introduce the angle $\alpha$ between
$\bB_e$ and $\bB_i$\/, so $\bB_e = B_e(\cos\alpha,\sin\alpha,0)$. Since the MHD equations are invariant under the substitution $-\bB$ for $\bB$\/, we  can always choose such the direction of vector $\bB_e$ that the angle $\alpha$ is either acute or right.  Hence, in what follows we assume that $0 \leq \alpha \leq \pi/2$.  Finally, we introduce the dimensionless parameters $\zeta = \rho_i/\rho_e$ and $\chi = B_i/B_e$\/. We rewrite Eq.~(\ref{eq:3.9}) as
\begin{equation} \omega^2 = \frac{g^2 h}{V_{Ae}^2} \frac{h\big[\chi^2\cos^2\varphi + \cos^2(\varphi -
\alpha)\big] +1 - \zeta} {\zeta + 1},  \label{eq:4.1} \end{equation}
where the Alfv\'en speed in the lower medium, $V_{Ae}$\/, and the dimensionless wave number $h$ are defined by  
\begin{equation} V_{Ae}^2 = \frac{B_e^2}{\mu_0\rho_e}, \quad 
h = \frac{V_{Ae}^2 k}g . \label{eq:4.2} \end{equation}
A well-known result is that the interface is stable when $\zeta < 1$, i.e.\
when the density of the upper medium is smaller than that of the lower medium.
In the opposite situation, i.e.\ when $\zeta > 1$, there is a qualitative
difference between the case where the magnetic field is in the same direction at
the two sides of the interface ($\alpha = 0$), and the case where the magnetic
field is sheared ($\alpha \neq 0$). In the first case perturbations with the
wave vector perpendicular to the magnetic field ($\varphi = \pi/2$) are unstable
for any value of the dimensionless wave number $h$\/. Since, for such
perturbations, the instability increment or growth rate is equal to 
$$ 
\sqrt{gk\frac{\zeta - 1}{\zeta + 1}}, 
$$ 
the instability growth rate is unbounded. This means that the initial value problem describing the evolution of the interface initial perturbation is ill-posed. Of course, the growth rate will be bounded and the problem will be well-posed if we take into account either dissipation or the finite thickness of the transition between the two homogeneous regions. 
\begin{figure} \centerline{\includegraphics[width=20pc]{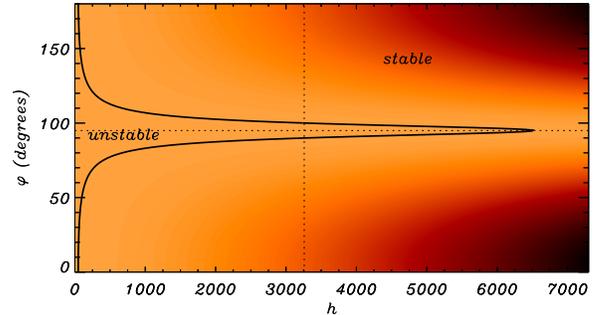}}
\caption{Square of the frequency (in arbitrary units) as a function of the
dimensionless number $h$ and $\varphi$ for the interface. The continuous curve corresponds to $\omega=0$ and represents the
transition between stable and  unstable modes. The horizontal and vertical
dotted lines, respectively, represent the critical angle $\varphi_c$\/, and the
wave number $\bar{h}_c/2$ at which the increment takes its maximum value. In this
plot, $\alpha=10^{\circ}$, $\chi=1$, and $\zeta=100$.} \label{w2kphi} 
\end{figure}

In the second case ($\alpha \neq 0$) perturbations with a fixed direction of the wave number defined by the angle $\varphi$ are unstable only when 
\begin{equation}
h < h_c(\varphi) = \frac{\zeta - 1}{\chi^2\cos^2\varphi + \cos^2(\varphi - \alpha)} .
\label{eq:4.3}
\end{equation}

In Fig.~\ref{w2kphi} we have plotted the square of $\omega$ as a function of $h$ and $\varphi$\/. The continuous curve corresponds to $\Omega=0$ and separates the regions between stable and unstable modes. In fact all perturbations with the wave number larger than $\bar{h}_c$ are stable, where

\begin{eqnarray}
\bar{h}_c &=& \max_\varphi h_c(\varphi) = h_c(\varphi_c) \nonumber\\
&=& {\frac{\zeta - 1}{2\chi^2\sin^2\alpha}}\left(\chi^2 + 1 + 
   \sqrt{\chi^4 + 2\chi^2\cos2\alpha + 1}\right) , \hphantom{xx}
\label{eq:4.4}
\end{eqnarray}
and 
\begin{equation}
\varphi_c = \frac12\arctan\frac{\sin2\alpha}{\chi^2 + \cos2\alpha} + \frac\pi2 .
\label{eq:4.5}
\end{equation}
Note that $\varphi_c$ is defined with the accuracy up to an additive constant
multiple to $\pi$\/. 

It turns out that the instability increment takes its maximum value $\gamma_m$
for a harmonic perturbation with the dimensionless wave number $\bar{h}_c/2$
(see vertical line in Fig.~\ref{w2kphi}) and  propagating at either the angle $\varphi_c$ (see horizontal line in Fig.~\ref{w2kphi}) or $\varphi_c + \pi$. This maximum value is given by 
\begin{equation}
\gamma_m = \frac g{2V_{Ae}}\sqrt{\bar{h}_c\frac{\zeta - 1}{\zeta + 1}} .
\label{eq:4.6}
\end{equation}
Hence, in the case of sheared magnetic field, the perturbation growth rate is
bounded, and the initial value problem is well-posed.

The external and internal Alfv\'en speed satisfy the following relationship
\begin{equation}
V_{Ae} = V_{Ai} \frac {\sqrt{\zeta}}{\chi},
\label{vaivvae}
\end{equation}
and now we rewrite Eq.~(\ref{eq:4.6}) in terms of the internal Alfv\'en speed,
\begin{equation}
\gamma_m = \frac{g|\zeta - 1|}{2\sqrt{2}V_{Ai}\sin \alpha}
\sqrt{\frac{\chi^2 + 1 + \sqrt{\chi^4 + 2\chi^2\cos2\alpha + 1}} 
{\zeta(\zeta + 1)}}.
\label{gaminterface}
\end{equation}
\begin{figure} \centerline{\includegraphics[width=20pc]{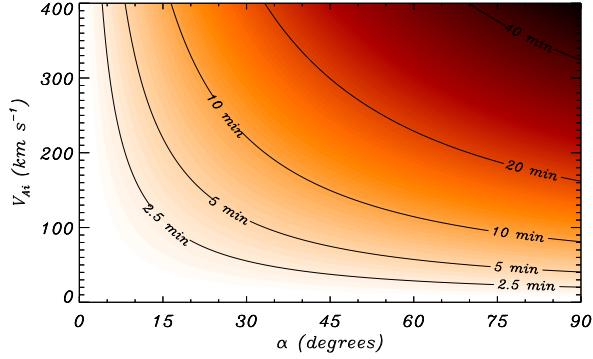}}
\caption{Growth time $\tau_g = 1/\gamma_m $ of the MRT unstable
mode at a single interface as a function of the shear angle $\alpha$ and
$V_{Ai}$ for $\chi=1$ and $\zeta=100$.} \label{tginterface} \end{figure}
It is interesting to study the dependence of the growth time, defined as $\tau_g = 1/\gamma_m$,  on the equilibrium parameters. In Fig.~\ref{tginterface} the two-dimensional dependence of $\tau_g$ is represented as a function of $\alpha$ and $\chi$ for $\zeta = 100$. We do see that a given growth time can be obtained by the proper combination of the parameters $\alpha$, $\chi$, and $V_{Ai}$. Note that, for small angles, a larger Alfv\'en speed is required to obtain the same growth rate. In fact, Eq.~(\ref{gaminterface}) is further simplified if we take the limit of small shear angles ($\alpha\ll1$),
\begin{equation}
\gamma_m \approx \frac{g|\zeta - 1|}{2\alpha V_{Ai}}
\sqrt{\frac{\chi^2 + 1}{\zeta(\zeta + 1)}}.
\label{gaminterfaceapprox}
\end{equation}
This expression explains why decreasing $\alpha$ while keeping $\gamma_m$ constant
requires an increase of the internal Alfv\'en speed (if the rest of the
parameters, i.e., $g$, $\chi$ and $\zeta$ are constant). In the opposite limit,
i.e., when $\alpha\approx \pi/2$, Eq.~(\ref{gaminterface}) reduces to
\begin{equation}
\gamma_m \approx \frac {g}{2 V_{Ai}}\frac{|\zeta - 1|}{\sqrt{\zeta\left({\zeta +
1}\right)} }\Theta.
\label{gaminterfaceapproxalphalarge}
\end{equation}
where
\begin{equation}
\Theta= \left\{\begin{array}{cl} 
\chi, & \chi>1, \vspace*{2mm} \\
1, & \chi \le 1.
\end{array}\right.
\label{eqchi}
\end{equation}
Thus, as Fig.~\ref{tginterface} indicates, for configurations  with a strong
shear the curve of the growth time is almost horizontal since, according to
Eq.~(\ref{gaminterfaceapproxalphalarge}), it is independent of $\alpha$. Note that for $\chi<1$ the growth time is independent of $\chi$ while it is linearly
proportional to $\chi$ for $\chi>1$. Finally note that the factor that contains the dependence with $\zeta$ in the previous expressions is approximately 1 in the limit of  $\zeta \gg 1$. Then
Eqs.~(\ref{gaminterface})-(\ref{gaminterfaceapproxalphalarge}) can be further
simplified for configurations with a high density contrast.

\subsection{Stability of a magnetic slab}
\label{subsec:stab_slab}

Since now there is a natural spatial scale $a$\/, it is convenient to introduce
a new dimensionless wave number $\kappa = ak$\/. We also introduce the parameter
characterising the relative strength of magnetic field and gravity, $\sigma =
V_{Ae}^2/(ag)$. Otherwise we use the same dimensionless parameters as in the
previous section. Then we rewrite Eq.~(\ref{eq:3.14}) as
\begin{equation}
\omega_\pm^2 = \frac{g(\tilde{F} \pm \tilde{G})}{a\tilde{H}} ,
\label{eq:4.7}
\end{equation}
where 
\begin{eqnarray}
\tilde{F} &=& \sigma\kappa^2\big\{\big[\chi^2\cos^2\varphi +
    \zeta\cos^2(\varphi-\alpha)\big]\cosh2\kappa \nonumber\\
&+& \big[\zeta\chi^2\cos^2\varphi + \cos^2(\varphi-\alpha)\big]\sinh2\kappa\} ,
\label{eq:4.8}
\end{eqnarray}
\begin{eqnarray}
\tilde{G} &=& \Big\{\sigma^2\kappa^4\big[\chi^2\cos^2\varphi -
    \zeta\cos^2(\varphi-\alpha)\big]^2 \nonumber\\
&+& \kappa^2(\zeta-1)^2\big[(\zeta^2 + 1)\sinh^22\kappa +
    \zeta\sinh4\kappa\big]\Big\}^{1/2} , \hphantom{xx} 
\label{eq:4.9}
\end{eqnarray}
\begin{equation}
\tilde{H} = 2\zeta\cosh2\kappa + (\zeta^2 + 1)\sinh2\kappa .
\label{eq:4.10}
\end{equation}

\noindent It is obvious that only $\omega_-^2$ can be negative, while
$\omega_+^2$ is always positive. Hence, only the sausage perturbations can be
unstable, while the kink perturbations are always stable. If $\omega_-^2 <
0$, then the two roots of Eq.~(\ref{eq:3.13}) considered as a quadratic equation
with respect to $\omega^2$ have different signs. This is only possible when the
free term of the quadratic Eq.~(\ref{eq:3.13}) is negative. In the dimensionless
variables this condition is written as 
\begin{eqnarray}
\kappa^2\sigma^2\big\{2\chi^2\cos^2\varphi\cos^2(\varphi-\alpha) &+&
   \big[\cos^4(\varphi-\alpha) \nonumber\\
+\,\chi^4\cos^4\varphi\big]\tanh2\kappa\big\} &<&
   (\zeta - 1)^2\tanh2\kappa . 
\label{eq:4.11}
\end{eqnarray}
This inequality can be rewritten as   
\begin{equation}
f(\kappa) \equiv \kappa(X + Y\tanh2\kappa]) - Z\frac{\tanh2\kappa}\kappa < 0 ,
\label{eq:4.12}
\end{equation}
where
\begin{equation}
\begin{array}{l}
X = 2\chi^2\cos^2\varphi\cos^2(\varphi-\alpha), \vspace*{1.5mm}\\
Y = \cos^4(\varphi-\alpha) + \chi^4\cos^4\varphi, \vspace*{1.5mm}\\
Z = \sigma^{-2}(\zeta - 1)^2 .
\end{array}
\label{eq:4.13}
\end{equation}
Differentiating the function $f(\kappa)$ we obtain
\begin{equation}
f'(\kappa) = X + Y\tanh2\kappa + \frac{2\kappa Y}{\cosh^22\kappa} +
   \frac{Z[\sinh(4\kappa) - 4\kappa]}{2\kappa^2\cosh^22\kappa} . 
\label{eq:4.14}
\end{equation}
It follows from a well-known inequality $\sinh x > x$ for $x > 0$ that the last term on the right-hand side of this equation is positive. Hence, $f'(\kappa) > 0$. We also have $f(\kappa) \to -2Z < 0$ as $\kappa \to 0$ and $f(\kappa) \to \infty$ as $\kappa \to \infty$\/. This implies that there is exactly one number $\kappa_c(\varphi)$ such that $f(\kappa_c) = 0$. The quantity $\kappa_c(\varphi)$ is defined by the equation obtained from Eq.~(\ref{eq:4.11}) by substituting the sign ``$<$'' by ``$=$''. The inequality~(\ref{eq:4.11}) is satisfied when  $\kappa < \kappa_c(\varphi)$, and
it is not satisfied otherwise. All perturbations with $\kappa > \bar{\kappa}_c =
\max\kappa_c(\varphi)$ are stable. In general, we failed to calculate
$\max\kappa_c(\varphi)$ analytically, so it must be done numerically.  

\begin{figure} \centerline{\includegraphics[width=20pc]{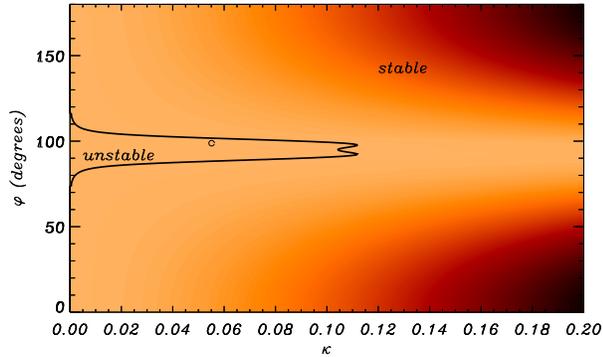}}
\caption{Square of the frequency (in arbitrary inits) as a function of $\kappa$ and $\varphi$ for the slab problem. The continuous curve corresponds to $\omega^2=0$ and represents the transition between stable and MRT unstable modes. The circle indicates the position in the diagram of the maximum growth rate. In this plot, $\alpha=10^{\circ}$, $\chi=1$, $\zeta=100$, $\sigma=3.65\times10^4$\/.}  \label{w2kphislab} \end{figure}

\begin{figure} \centerline{\includegraphics[width=20pc]{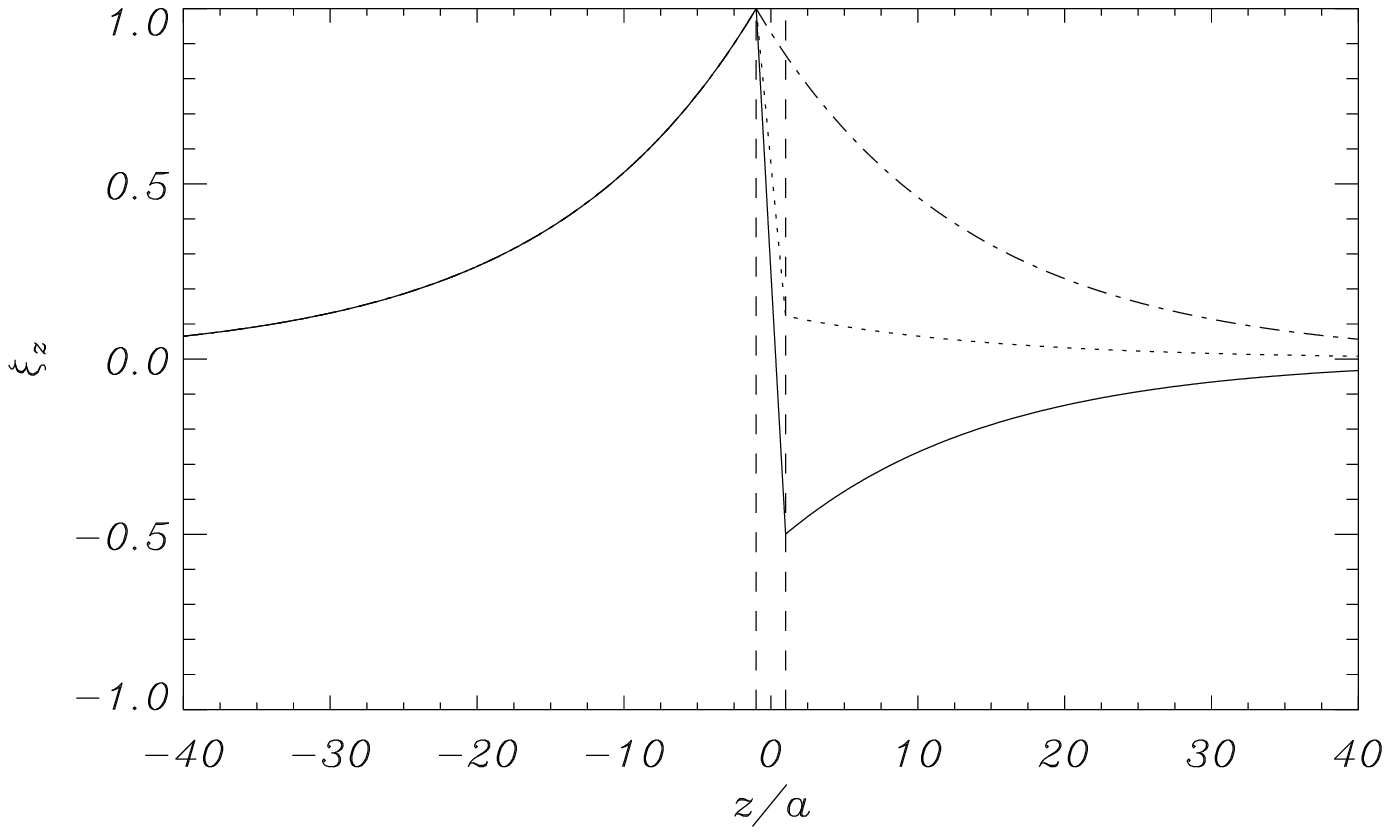}}
\centerline{\includegraphics[width=20pc]{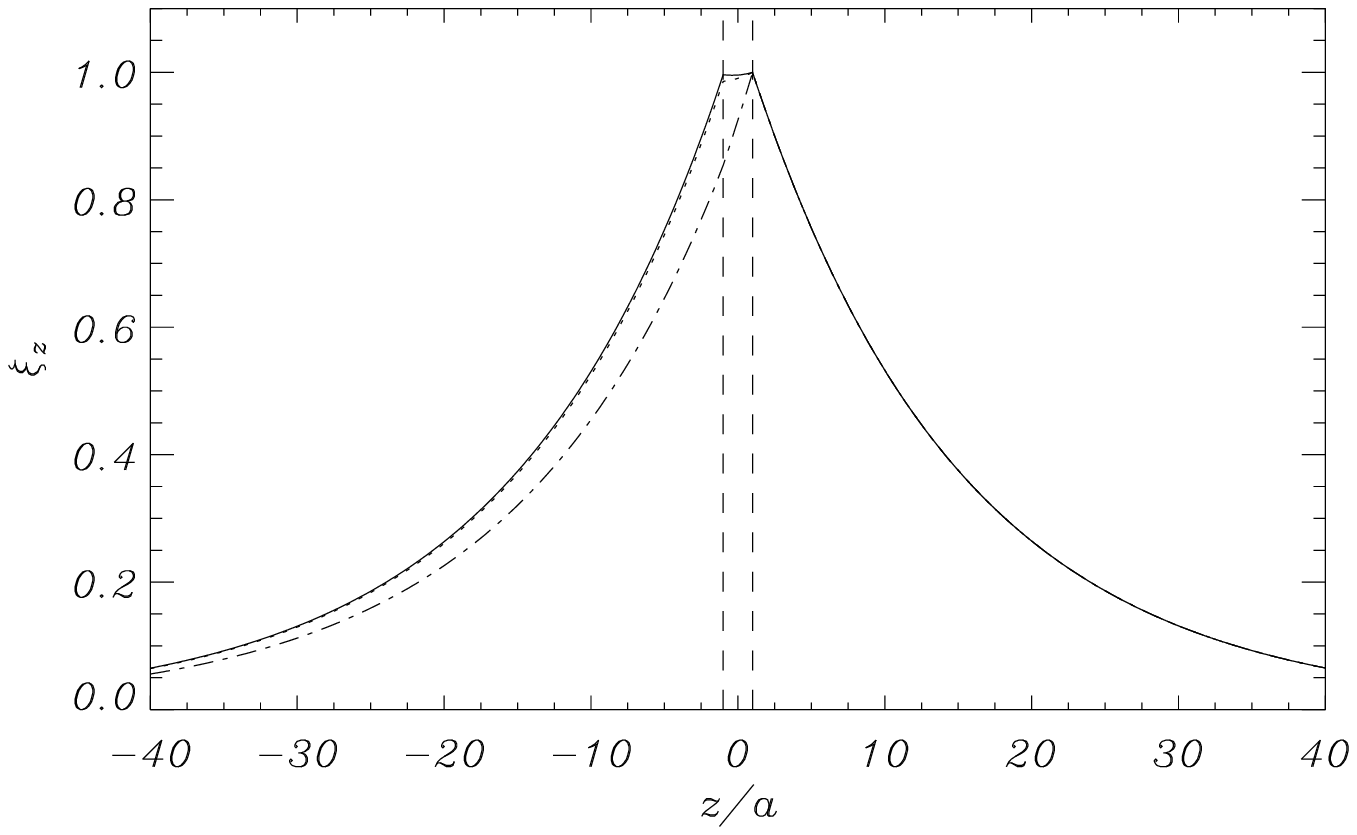}} 
\caption{Eigenfunction for the $\omega_-^2$ solution (top panel) and for the
stable $\omega_+^2$ solution (bottom panel) in the slab model. The continuous
curve corresponds to an angle of propagation $\varphi=80^{\circ}$, the
dotted curve to $\varphi=90^{\circ}$, and the dashed-dot to
$\varphi=100^{\circ}$. A fixed dimensionless wave number $\kappa = 0.07$ has been selected to calculate the eigenfunctions. The rest of the parameters are the same as in Fig.~\ref{w2kphislab}. The vertical dashed lines represent the slab boundaries. Note that  some parts of the dotted and dashed-dotted lines are superimposed on the solid lines and therefore not easy to distinguish. This happens on the left part of the top panel and on the right part of the bottom panel.} \label{eigens} \end{figure}

An example of the dependence of the square of the frequency on $\kappa$ and
$\varphi$ is shown in Fig.~\ref{w2kphislab}. The value of $\sigma=3.65\times10^4$ used to plot this figure can be obtained if we take, for example, $V_{Ae} = 10^3$~km\,s$^{-1}$ and $a = 100$~km. It is interesting to compare this figure with the results for the single interface shown in Fig.~\ref{w2kphi}. The equilibrium parameters are exactly the same in the two plots, but now for the slab problem we have an additional parameter, which is the half width of the slab, denoted by $a$\/. The curve representing the transition between the stable and the unstable regime of the $\omega_-^2$ solution is slightly more complex for the slab problem and shows a double lobe structure around the maximum $\kappa$. The eigenfunctions for three different propagation angles are plotted in Fig.~\ref{eigens} for a fixed $\kappa$ and the same parameters as in Fig.~\ref{w2kphislab}. In the top panel the $\omega_-^2$ solution changes from stable (continuous curve) with a clear ``sausage" character to unstable (dotted and dashed-dotted lines). Note that eventually the eigenfunction of this mode is now localised at the lower interface. On the contrary, the solution corresponding to the $\omega_+^2$, see bottom panel of Fig.~\ref{eigens}, has a clear ``kink" profile, but it tends to be localised at the upper interface when $\varphi$ is increased. Therefore the nature of the modes is closer to surface waves associated to the individual interfaces. Similar results were found in \citet{terretal2012} in the absence of magnetic shear.

\begin{figure} \centerline{\includegraphics[width=20pc]{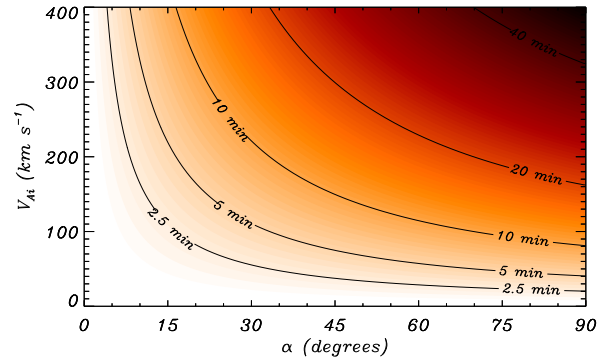}}
\caption{Growth time of the MRT instability for the slab problem as a function of the shear angle $\alpha$ and $V_{Ai}$. In this plot $\chi=1$, $\zeta=100$ and $a=100\,\rm km$.}
\label{tgslab} \end{figure}

We concentrate now on the analysis of the growth time. In Fig.~\ref{tgslab} the
dependence of $\tau_g$ associated to the most unstable mode is plotted as a function of $\alpha$ and $V_{Ai}$ for a fixed value of $\chi$, $\zeta$, and $a$. The differences with respect to the interface results, see Fig.~\ref{tginterface}, are that the curves of constant $\tau_g$ have essentially shifted down in the diagram. This means that the slab configuration is more stable than the interface model. This diagram can be used as a diagnostic tool and clearly shows the dependence of the grow times in the space of parameters.

The stability analysis is greatly simplified in two limiting cases. In the fist
case the magnetic fields in the slab and external regions are almost parallel,
$\alpha \ll 1$, and $|\zeta-1|/\sigma$ is of the order of unity. In this case it
follows from the equation $f(\kappa) = 0$ that $\kappa_c(\varphi)$ takes
moderate values when $\varphi$ is not close to $\pi/2$, while it takes very
large values when $\varphi$ is close to $\pi/2$\/. Hence, to calculate
$\bar{\kappa}_c$\/, it is enough to consider $\varphi$ close to $\pi/2$\/. In
accordance with this we put $\varphi = \pi/2 - \psi$ and assume that $|\psi| \ll
1$. In addition, since $\kappa_c \gg 1$, we can take $\tanh2\kappa_c \approx 1$.
Then we obtain from the equation $f(\kappa) = 0$ the approximate expression 

\begin{equation}
\kappa_c(\varphi) \approx \frac{|\zeta - 1|}{\sigma[(\psi + \alpha)^2 + \chi^2\psi^2]} . 
\label{eq:4.15}
\end{equation}
It immediately follows that
\begin{equation}
\bar{\kappa}_c = \kappa_c(\varphi_c) \approx \frac{|\zeta - 1|(1 + \chi^2)}
   {\sigma\chi^2\alpha^2} , \quad
\varphi_c \approx \frac\pi2 + \frac\alpha{1 + \chi^2} .
\label{eq:4.16}
\end{equation}
It is also not difficult to obtain the asymptotic expression for $\omega_-^2$ valid for $\alpha \ll 1$ and $|\varphi - \pi/2| \ll 1$. It is given by
\begin{equation} 
\omega_-^2 \approx \frac{g\{\sigma\kappa^2[\chi^2\psi^2 + (\psi + \alpha)^2] -
   \kappa|\zeta - 1|\}}{a(\zeta + 1)}.
\label{eq:4.17}
\end{equation}
When deriving this expression we have assumed that $\kappa \gg 1$ because
$|\omega_-^2|$ is of the order of $g/a$ when $\kappa \sim 1$, while
$|\omega_-^2| \gg g/a$ when $\kappa \gg 1$. The quantity $\omega_-^2$ takes its
minimum value at $\varphi = \varphi_c$ and $\kappa = \frac12\bar{\kappa}_c$\/,
and the maximum growth rate is given by

\begin{equation} 
\gamma_m \approx \frac{|\zeta - 1|}{2\chi\alpha}\sqrt{\frac{g(1 + \chi^2)}{a\sigma(\zeta + 1)}}.
\label{eq:4.18}
\end{equation}
In terms of the internal Alfv\'en speed the previous expression reduces to
\begin{equation}
\gamma_m \approx \frac{g|\zeta - 1|}{2\alpha V_{Ai}}
\sqrt{\frac{\chi^2 + 1}{\zeta(\zeta + 1)}}.
\label{gamslabeapprox}
\end{equation}
This is exactly the same as Eq.~(\ref{gaminterfaceapprox}) which corresponds to
the interface result in the limit of small $\alpha$. This confirms that the slab
problem in the limit of $\kappa \gg 1$ reduces to the interface problem because 
the penetration scale given by $1/\kappa$ is rather small and the role of the
upper interface is negligible for the unstable mode (see also the plot of the
eigenfunctions in Fig.~\ref{eigens}). Again we see that the situation here is similar to one that we have in the case of a single interface: $\gamma_m \to \infty$ as $\alpha \to 0$, so, in the case of parallel magnetic field ($\alpha = 0$), the growth rate is unbounded and the initial value problem is ill-posed. On the other hand,  the growth rate is bounded when the magnetic field is sheared ($\alpha \neq 0$), so  the initial value problem is well-posed. As mentioned in the case of a single interface, the growth rate of the MRT instability in the model with parallel magnetic field ($\alpha = 0$) will be bounded and the problem will be well-posed if we take into account either dissipation or the finite thickness of the transitions between the three homogeneous regions.

Another limiting case where the analytic asymptotic analysis is possible is when
$\zeta \gg 1$ and $\sigma \gtrsim \zeta^2$\/, while $\chi \simeq 1$ and $\alpha \simeq 1$. In this case it is not difficult to see from the equation $f(\kappa) = 0$ that $\kappa_c(\varphi) \ll 1$, so the equation $f(\kappa) = 0$ can be written in the approximate form as
\begin{eqnarray}
\kappa^2\big[\cos^4(\varphi &-& \alpha) + \chi^4\cos^4\varphi\big] \nonumber\\
&+& \kappa\chi^2\cos^2\varphi\cos^2(\varphi-\alpha) =  \zeta^2\sigma^{-2} .
\label{eq:4.19}
\end{eqnarray}
This equation is used in Appendix~\ref{app:A} to calculate $\varphi_c$\/. It is found that 
\begin{equation} 
\begin{array}{l} \displaystyle
\varphi_c = \varphi_{c1} \approx \frac\pi2 + 
   \frac{2\zeta\cot\alpha}{\sigma\chi^2}, \vspace*{2mm}\\
\displaystyle \bar{\kappa}_c = \kappa_c(\varphi_{c1}) \approx
   \frac\zeta{\sigma\sin^2\alpha}
\end{array}
\label{eq:4.20}
\end{equation}
when $\chi > 1$ and 
\begin{equation}
\begin{array}{l} \displaystyle
\varphi_c = \varphi_{c2} \approx \frac\pi2 + \alpha - 
   \frac{2\zeta\cot\alpha}\sigma, \vspace*{2mm}\\
\displaystyle \bar{\kappa}_c = \kappa_c(\varphi_{c2}) \approx
   \frac\zeta{\sigma\chi^2\sin^2\alpha}
\end{array}  
\label{eq:4.21}
\end{equation}
when $\chi < 1$.

\begin{figure}
\centerline{\includegraphics[width=20pc]{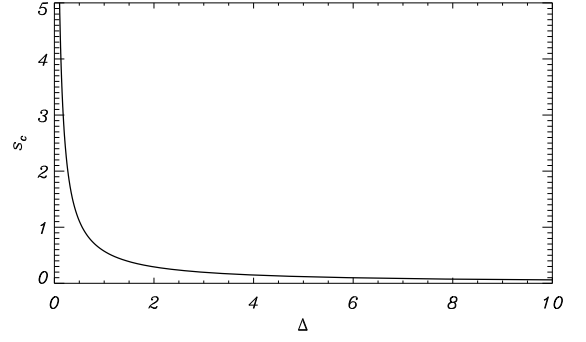}} 
\caption{The dependence of $s_c$ on $\Delta$\/.}
\label{fig:s_Delta}
\end{figure}
\begin{figure}
\centerline{\includegraphics[width=20pc]{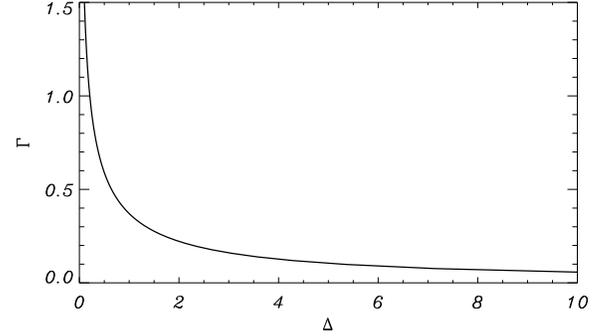}}
\caption{The dependence of $\Gamma$ on $\Delta$\/.}
\label{fig:gamma_m}
\end{figure}

It is shown in Appendix~\ref{app:B} that the fastest growing mode propagates approximately perpendicular to the external magnetic field ($\varphi \approx \pi/2 + \alpha$). Its dimensionless wave number is equal to $\kappa_c = \zeta^{-1}s_c(\Delta)$, where
\begin{equation}
\Delta = \frac{\sigma\chi^2\sin^2\alpha}{\zeta^2} .
\label{eq:4.22}
\end{equation}
The dependence of $s_c$ on $\Delta$ is shown in Fig.~\ref{fig:s_Delta}. The instability increment is equal to
\begin{equation}
\gamma_m = (g/a\zeta)^{1/2}\Gamma(\Delta). \label{gammaslabapprox}
\end{equation}
The dependence of $\Gamma$ on $\Delta$ is plotted in Fig.~\ref{fig:gamma_m}. All these results are obtained under the assumption that $\zeta\Delta > 1$, i.e., when $\chi^2\sin^2\alpha > \zeta/\sigma$. 

\begin{figure} \centerline{\includegraphics[width=20pc]{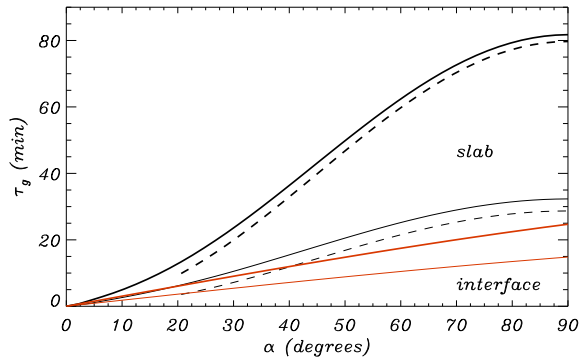}}
\caption{Growth time of the MRT instability for the slab problem (black colour) and interface problem (red colour) as a function of the shear angle $\alpha$. The thin curves correspond to $\sigma = 1.45\times10^5$ ($V_{Ai}=200\rm\,km\, s^{-1}$) while thick curves represent the case $\sigma = 5.2\times10^4$ ($V_{Ai} = 120\rm\,km\, s^{-1}$). Dashed lines correspond to the approximation in the slab problem for large angles given by Eq.~(\ref{gammaslabapprox}). In this plot
$\chi=1$ and $\zeta=100$.}  \label{tgslab1}
\end{figure}

In Fig.~\ref{tgslab1} the growth time of the instability
calculated using the full solution (continuous curve) is plotted together with
the approximation for finite angles (dashed curves) given by
Eq.~(\ref{gammaslabapprox}). We do see a good match in the behaviour of the two
curves. The approximation slightly underestimates the growth time, and the
differences increase when $\sigma$ decreases. This is what we
should expect if we recall that the approximation is based on the assumption
$\sigma \gg 1$. We have already shown that, for small angles of the shear, the
interface and the slab results are the same. Now this is also
evident in Fig.~\ref{tgslab1}. This plot also shows that the differences in the
growth time between the slab and the interface can be quite significant for
$\alpha\simeq 1$.

\section{Application to oscillating threads}

The purpose of this section is to use the previous theoretical results to  infer some information about the shear in real prominence threads. We consider the thread oscillation studied by \citet{okamoto07} using HINODE. These authors found threads oscillating vertically with periods around 4 min. \citet{TERetal2008} used these oscillations to do a seismological study of the thread. Although \citet{okamoto07} did not investigate the lifetime of the threads, from the movie of the event we estimate that threads do not last long. It is found that the typical lifetime is 10 min. The main idea here is to assume that the excitation of nstable modes is responsible for the short lifetime of the structure. This MRT instability produces the disappearance of the structure. According to our results the fastest growing mode when there is magnetic shear has the following approximate growth time (see
Eq.~(\ref{gaminterfaceapprox}))
\begin{equation} \tau_g \approx
\frac{\sqrt{2}}{g}V_{Ai}\alpha, \label{growthintersimp} 
\end{equation}
where we have assumed that $\zeta \gg  1$ and $\chi \approx 1$. This expression
applies to the interface problem as well as to the slab problem (see
Eq.~(\ref{gamslabeapprox})), and it is valid in the limit of small $\alpha$ only. If the growth time is assumed to be around $10\rm \,min$ in order to use
Eq.~(\ref{growthintersimp}) we must obtain a small angle, otherwise we get an
inconsistent result. \citet{TERetal2008} found that the lower bound for the internal Alfv\'en speed is between 120 and $350\,\rm km\, s^{-1}$. Nevertheless, some of the threads have high Alfv\'en velocities, up to $800\,\rm km s^{-1}$, since they belong to an active region prominence. Let us assume that the velocity is around $500 \,\rm km\, s^{-1}$ and calculate the corresponding shear angle. According to Eq.~(\ref{growthintersimp}) and using that $\tau_g=10\,\rm min$ and $V_{Ai}=500\,\rm km\,s^{-1}$ we obtain that $\alpha\approx 13^{\circ}$. This angle is small and the application of Eq.~(\ref{growthintersimp}) is justified. This is in agreement with the behaviour found in the more general expression given by Eq.~(\ref{gaminterface}) which shows that $\alpha$ decreases when the internal Alfv\'en speed increases  (see also Fig.~\ref{tginterface}). In fact this expression shows that $\alpha$ is small for sufficiently large values of $V_{Ai}$ but moderate for smaller values. For the present case of a velocity of $V_{Ai}=500\,\rm km\,s^{-1}$ the angle is small, and this is in accordance with the observations of prominence threads.

The observed oscillations of the thread are most probably kink oscillations. It is well known that the sausage waves in magnetic slabs and and tubes have quite similar properties, while the properties of kink waves in slabs are quite different from those in tubes (e.g.\ \citealp{Edwin1982,Edwin1983}). Since $\omega_+$\/-mode is a kink mode, it is quite improbable that the observed oscillations of the thread are described by this mode. Note that, in the linear approximation, the kink oscillation of the thread does not interact with the unstable mode causing its disappearance, and thus it does not affect the instability growth time.      

\section{Summary and conclusions} \label{sec:sum}

In the present paper we have extended the study by \citet{terretal2012} to the
models of an interface and a slab with magnetic shear, and have focused on the
MRT instability. The fact that the magnetic field changes its direction
introduces a bounded growth rate of the instability. This is different from the
models without shear where the growth rate is unbounded. Because of this
\citet{terretal2012} concentrated on a particular wave number in the
perpendicular direction equal to $1/a$\/, $a$ being the half-thickness of the
slab. Here we have focused on the maximum growth rate, representing the most
unstable mode of the system, and have found analytical expressions in various
limiting cases. For small angles of the shear the growth time is linearly
proportional to the shear angle $\alpha$\/. This applies to the single interface
as well as to the slab problem. On the contrary, for large angles the growth
time depends only weakly on $\alpha$ in the interface problem. In this limit,
we have also been able to calculate an approximate expression for the growth
time in the slab problem. 

We have shown, using a simple example, how it is possible to estimate the shear
angle in threads belonging to active region prominences from the combination of
observations and the theoretical results presented in this paper. Using this
method we have found that the observations of oscillating threads of
\citet{okamoto07} are compatible with small shear angles (around $13^{\circ}$).
This indirect method of inferring some of the equilibrium properties of threads
can be potentially used as a seismological tool.

Several simplifications have been done in the models considered in this work.
Both the interface and slab configurations are unbounded in the perpendicular
direction ($y$\/-direction) and we have assumed that there is dense material along
the full length of the magnetic tube. However, in reality, threads only
represent a small part of the full magnetic tube. Compressibility
and partial ionisation \citep[see][]{diaz2012} have been also ignored. These
basic assumptions in the models have enabled us to derive analytical expressions
for the growth times. Nevertheless, the need for using improved models is
obvious, and the study of the nonlinear evolution of the system is very relevant
to asses the role of the MRT instability in the fast disappearance of threads. \vspace*{5mm}

\noindent 
{\bf Acknowledgements}. A part of this work was carried out when MSR was a guest of Departament de F\'{\i}sica of Universitat de les Illes Balears. He acknowledges the financial support received from the Universitat de les Illes Balears and the warm hospitality of the Departament. He also acknowledges the support by the
STFC grant. J.T. acknowledges support from the Spanish Ministerio de Educaci\'on
y Ciencia through a Ram\'on y Cajal grant and financial support from
MICINN/MINECO and FEDER Funds through grant AYA2011-22846 Funding from CAIB
through the Grups Competitius scheme and FEDER Funds is also acknowledged.

\appendix
\section{Calculation of $\varphi_c$ and $\bar{\kappa}_c$ for the slab model.}
\label{app:A}

In this appendix we calculate $\varphi_c$ and $\bar{\kappa}_c$ for the slab
model in the case where $\sigma \gg \zeta \gg 1$, while $\chi \simeq 1$ and
$\alpha \simeq 1$. As it is shown in Sect.~\ref{subsec:stab_slab}, in this case
$\kappa_c(\varphi)$ is defined by the approximate Eq.~(\ref{eq:4.19}). To
calculate $\varphi_c$ we differentiate Eq.~(\ref{eq:4.19}) with respect to
$\varphi$ and then take $d\kappa/d\varphi = 0$. As a result we obtain

\begin{eqnarray}
2\kappa\big[\chi^4\cos^3\varphi\sin\varphi &+& 
   \cos^3(\varphi-\alpha)\sin(\varphi-\alpha)\big] \nonumber\\
&+& \chi^2\cos\varphi\cos(\varphi-\alpha)\sin(2\varphi-\alpha)  = 0 .
\label{eq:A1}
\end{eqnarray}
Eliminating $\kappa$ from this equation and Eq.~(\ref{eq:4.19}) we obtain the
equation for $\varphi_c$\/:
\begin{eqnarray}
\sin\alpha\cos^2\varphi\cos^2(\varphi-\alpha)\sin(2\varphi-\alpha)
  \big[\cos^4(\varphi \!\!&-&\!\!  \alpha) - \chi^4\cos^4\varphi\big] \nonumber\\
= \zeta^2\sigma^{-2}\chi^{-4}\big\{\cos^2(\varphi-\alpha)
  \sin[2(\varphi-\alpha)]\!\!&+&\!\! \chi^4\cos^2\varphi\sin2\varphi\}^2 .
\label{eq:A2}
\end{eqnarray}
Since $\zeta^2\sigma^{-2} \ll 1$, to obtain the solution to this equation we use
the regular perturbation method. In the first-order approximation we obtain that
the left-hand side of Eq.~(\ref{eq:A2}) is zero. It is possible when one of
the four multiplier that depend on $\varphi$ is zero. We investigate these
possibilities separately.

(i) Let $\cos\varphi = 0$, i.e.\ $\varphi = \pi/2$. In the second-order
approximation we look for the solution in the form $\varphi = \pi/2 - \psi$\/,
where $|\psi| \ll 1$. Substituting this expression in Eq.~(\ref{eq:A2}) we easily calculate $\psi$ and eventually obtain
\begin{equation} 
\varphi_{1\pm} \approx \frac\pi2 \pm \frac{2\zeta\cot\alpha}{\sigma\chi^2}.
\label{eq:A3}
\end{equation}

(ii) Let $\cos(\varphi-\alpha) = 0$, i.e.\ $\varphi = \pi/2 + \alpha$\/. In the second-order approximation we look for the solution in the form $\varphi = \pi/2 + \alpha - \psi$\/, where 
$|\psi| \ll 1$. Substituting this expression in Eq.~(\ref{eq:A2}) we, once again, easily calculate $\psi$ and obtain
\begin{equation} 
\varphi_{2\pm} \approx \frac\pi2 + \alpha \pm \frac{2\zeta\cot\alpha}\sigma .
\label{eq:A4}
\end{equation}

(iii) Let $\sin(2\varphi-\alpha) = 0$, i.e.\ $\varphi = \frac12\alpha$ or $\varphi = (\pi + \alpha)/2$. In the second-order approximation we look for the solution in the form $\varphi = \frac12\alpha + \psi$\/, where $|\psi| \ll 1$. Substituting this expression in Eq.~(\ref{eq:A2}) we calculate $\psi$ and obtain
\begin{equation} 
\varphi_3 \approx \frac\alpha2 + \frac{2\zeta^2(1-\chi^4)}{\sigma^2\chi^4}
   \tan\frac\alpha2\,{\rm sec}^3\frac\alpha2 .
\label{eq:A5}
\end{equation}
Similarly, looking for the solution in the form $\varphi = (\pi + \alpha)/2 + \psi$ with $|\psi| \ll 1$, we obtain
\begin{equation} 
\varphi_4 \approx \frac\pi2\ + \frac\alpha2 + \frac{2\zeta^2(\chi^4-1)}
   {\sigma^2\chi^4}\cot\frac\alpha2\,{\rm cosec}^3\frac\alpha2 .
\label{eq:A5a}
\end{equation}

(iv) Let $\cos(\varphi-\alpha) = \pm\chi\cos\varphi$\/. It follows from this equation that
\begin{equation} 
\varphi = \varphi_0 \equiv \arctan\frac{\pm\chi - \cos\alpha}{\sin\alpha} .
\label{eq:A6}
\end{equation}
In the second-order approximation we look for the solution in the form $\varphi
= \varphi_0 + \psi$\/, where $|\psi| \ll 1$. Substituting this expression in
Eq.~(\ref{eq:A2}), after some algebra, we obtain in the second-order approximation
\begin{eqnarray} 
\psi\sin\alpha\sin(2\varphi_0 \!\!&-&\!\! \alpha)[\pm\chi\sin\varphi_0 - 
   \sin(\varphi_0 -\alpha)] \nonumber\\ 
&=&\! \zeta^2\sigma^{-2}\chi^{-6}[\tan\varphi_0(\cos\alpha \pm \chi) - \sin\alpha]^2 .
\label{eq:A7}
\end{eqnarray}
Using Eq.~(\ref{eq:A6}) we derive the formulae
\begin{equation}
\begin{array}{c} \displaystyle
\sin(2\varphi_0 - \alpha) = \frac{\sin\alpha(\chi^2 - 1)}
   {1 \mp 2\chi\cos\alpha + \chi^2}, \vspace*{2mm}\\  
\pm\chi\sin\varphi_0 - \sin(\varphi_0 -\alpha) = 
   \sqrt{1 \mp 2\chi\cos\alpha + \chi^2}.
\end{array}
\label{eq:A8}
\end{equation}
With the aid of Eqs.~(\ref{eq:A6}) and Eq.~(\ref{eq:A8}) we calculate $\psi$
from Eq.~(\ref{eq:A7}). Finally we obtain
\begin{equation} 
\varphi_{5\pm} \approx \arctan\frac{\pm\chi - \cos\alpha}{\sin\alpha} +
   \frac{\zeta^2(\chi^2-1)\sqrt{1 \mp 2\chi\cos\alpha + \chi^2}}
   {\sigma^2\chi^6\sin^4\alpha} .
\label{eq:A9}
\end{equation}
Hence, we have eight values of $\varphi$ that are the solutions of Eq.~(\ref{eq:4.19}), and we have to choose one at that $\kappa_c(\varphi)$ takes its maximum value. To do this we have to calculate $\kappa_c(\varphi)$ at each of these eight values of $\varphi$ using Eq.~(\ref{eq:A1}). The calculation is lengthy but straightforward, so we give only the final results:
\begin{equation} 
\begin{array}{c} \displaystyle
\kappa_c(\varphi_{1\pm}) \approx \frac{\pm\zeta}{\sigma\sin^2\alpha}, \quad
\kappa_c(\varphi_{2\pm}) \approx \frac{\mp\zeta}{\sigma\chi^2\sin^2\alpha}, 
   \vspace*{2mm} \\
\displaystyle \kappa_c(\varphi_3) \approx \frac{2\zeta^2}{\sigma^2\chi^2}\,\mbox{sec}^5\frac\alpha2, \vspace*{2mm} \\ 
\displaystyle \kappa_c(\varphi_4) \approx \frac{2\zeta^2}{\sigma^2\chi^2}\,\mbox{cosec}^5\frac\alpha2, \quad \kappa_c(\varphi_{5\pm}) \approx -\frac12 .  
\end{array}
\label{eq:A10}
\end{equation}
We disregard negative values of $\kappa_c$ because we assumed from the very beginning that $\kappa > 0$. Since we have assumed that $\zeta/\sigma \ll 1$, it is straightforward to see that $\kappa_c(\varphi_{1+}) > \kappa_c(\varphi_3),\,\kappa_c(\varphi_4)$ and $\kappa_c(\varphi_{2-}) > \kappa_c(\varphi_3),\,\kappa_c(\varphi_4)$. Hence, eventually,
\begin{equation} 
\bar{\kappa}_c \approx \kappa_c(\varphi_{1+})  
     \approx \frac\zeta{\sigma\sin^2\alpha}, \quad \chi > 1 ,
\label{eq:A11}
\end{equation}
and
\begin{equation} 
\bar{\kappa}_c \approx \kappa_c(\varphi_{2-}) 
     \approx \frac\zeta{\sigma\chi^2\sin^2\alpha}, \quad \chi < 1 .
\label{eq:A12}
\end{equation}

\section{Calculation of maximum increment.}
\label{app:B}

In this this section we calculate the maximum growth rate of the Rayleigh-Taylor
instability. In accordance with Eq.~(\ref{eq:A1}) the dimensionless wave number of an unstable mode is always small, $\kappa \lesssim \zeta/\sigma \ll 1$. This observation enables us to use the approximate Taylor expansions with respect to $\kappa$ for functions $\tilde{F}$\/, $\tilde{G}$ and $\tilde{H}$\/:
\begin{eqnarray}
\tilde{F} &=& \sigma\kappa^2\big\{\big[\chi^2\cos^2\varphi +
    \zeta\cos^2(\varphi-\alpha)\big]\big(1 + 2\kappa^2\big) \nonumber\\
&+& 2\kappa\big[\zeta\chi^2\cos^2\varphi + \cos^2(\varphi-\alpha)\big]\big\} ,
\label{eq:B1}
\end{eqnarray}
\begin{equation}
\tilde{G} = \Big\{\sigma^2\kappa^4\big[\chi^2\cos^2\varphi -
    \zeta\cos^2(\varphi-\alpha)\big]^2 + 
    4\kappa^3\zeta^3(1 + \zeta\kappa)\Big\}^{1/2} ,  
\label{eq:B2}
\end{equation}
\begin{equation}
\tilde{H} = 2\zeta(1 + \zeta\kappa) .   
\label{eq:B3}
\end{equation}
We introduce the dimensionless increment $\tilde{\gamma} =
(g/a)^{1/2}|\omega_-|$. Then we obtain from Eq.~(\ref{eq:4.7}) the approximate equation
\begin{equation}
\tilde{\gamma}^2 = 2\kappa^3\frac{\zeta^2 - 
   \sigma^2\kappa\{\kappa[\chi^4\cos^4\varphi + \cos^4(\varphi-\alpha)] +
   \chi^2\cos^2\varphi\cos^2(\varphi-\alpha)\}}{\tilde{F} + \tilde{G}} .
\label{eq:B4}
\end{equation}
Now we investigate this expression in three various intervals of variation of the angle $\varphi$\/.

(i) Let $\cos^2\varphi \simeq \cos^2(\varphi-\alpha) \simeq 1$, i.e.\ $\varphi$ be
not close to $\pi/2$ and to $\pi/2 + \alpha$\/. Then it follow from Eq.~(\ref{eq:B4}) that $\tilde{\gamma}^2 > 0$ only when $\kappa \lesssim \zeta^2/\sigma^2$\/. For these values of $\kappa$ we have $\tilde{F} + \tilde{G} \gtrsim \zeta^5/\sigma^3$ and $\tilde{\gamma} \lesssim (\zeta/\sigma)^{3/2}$\/.

(ii) Let $\varphi$ be close to $\pi/2$\/, so we take $\varphi = \pi/2 - \tilde\varphi$\/. Then we can use the approximate expression
\begin{equation}
\tilde{\gamma}^2 \approx 2\kappa^2\frac{\zeta^2 - \sigma^2\kappa\sin^2\alpha
   (\kappa\sin^2\alpha + \chi^2\tilde\varphi^2)}{\sigma\zeta\kappa\sin^2\alpha +
   \zeta(\sigma^2\kappa^2\sin^2\alpha + 4\zeta\kappa)^{1/2}} ,
\label{eq:B5}
\end{equation}
It is obvious that, for any fixed $\kappa$\/, $\tilde{\gamma}^2$ takes its
maximum value at $\tilde\varphi = 0$. When $\kappa \lesssim \zeta/\sigma^2$\/,
we obtain $\tilde{\gamma}^2 \simeq \zeta^2/\sigma^3 \ll \zeta^2/\sigma^2$\/. On
the other hand, we obtain $\tilde{\gamma}^2 \simeq \zeta^2/\sigma^2$ when $\kappa \simeq \zeta/\sigma$\/. Hence, when looking for the maximum value of
$\tilde{\gamma}$\/, we can take $\tilde\varphi = 0$ and $\kappa \simeq
\zeta/\sigma$\/. In that case $\sigma^2\kappa^2\sin^2\alpha \gg 4\zeta\kappa$
and we can further reduce Eq.~(\ref{eq:B5}) to
\begin{equation}
\tilde{\gamma}^2 \approx \frac{\kappa(\zeta^2 - \sigma^2\kappa^2\sin^4\alpha)}
   {\sigma\zeta\sin^2\alpha} .
\label{eq:B6}
\end{equation}
Then we easily find that the maximum value of $\tilde{\gamma}$ is given by
\begin{equation}
\tilde{\gamma}_m \approx \frac\zeta{\sigma\sin^2\alpha}\sqrt[4]{\frac4{27}},
\label{eq:B7}
\end{equation}
and it is taken at 
\begin{equation} 
\kappa \approx \frac\zeta{\sqrt3\sigma\sin^2\alpha} .
\label{eq:B8}
\end{equation}

(iii) Let now $\varphi$ be close to $\pi/2 + \alpha$\/, so we take $\varphi = \pi/2 + \alpha - \tilde\varphi$\/. Then we can use the approximate expression 
\begin{equation}
\tilde{\gamma}^2 \approx 2\kappa^3\frac{\zeta^2 - \sigma^2\kappa\chi^2\sin^2\alpha
   (\kappa\chi^2\sin^2\alpha + \tilde\varphi^2)}{\tilde F + \tilde G} ,
\label{eq:B9}
\end{equation}
where $\tilde F$ and $\tilde G$ are given by
\begin{equation}
\tilde F \approx \sigma\kappa^2\left[\chi^2\sin^2\alpha(1 + 2\zeta\kappa) + 
   \zeta\tilde\varphi^2\right],
\label{eq:B10}
\end{equation}
\begin{equation} 
\tilde F \approx \left[\sigma^2\kappa^4(\chi^2\sin^2\alpha -  \zeta\tilde\varphi^2)^2
   + 4\zeta^3\kappa^3(1 + \zeta\kappa)\right]^{1/2} .
\label{eq:B11}
\end{equation}
It is easy to show that $\tilde F + \tilde G$ is a monotonically increasing
function of $\tilde\varphi^2$\/. Since the numerator in Eq.~(\ref{eq:B9}) is a monotonically decreasing function of $\tilde\varphi^2$\/, we conclude that, at a fixed $\kappa$\/, $\tilde{\gamma}$ takes its maximum value at $\tilde\varphi = 0$\/. Hence, we can take
\begin{equation}
\tilde{\gamma}^2 \approx \frac{2\kappa^2(\zeta^2 - \sigma^2\kappa^2\chi^2\sin^4\alpha)}
   {\sigma\kappa\chi^2\sin^2\alpha(1 + 2\zeta\kappa) +
   [\sigma^2\kappa^2\chi^4\sin^4\alpha + 4\kappa\zeta^3(1 + \zeta\kappa)]^{1/2}} 
\label{eq:B12}
\end{equation}
when looking for the maximum value of $\tilde{\gamma}$\/. Introducing the new dimensionless variables
\begin{equation}
s = \zeta\kappa, \quad \Delta = \frac{\sigma\chi^2\sin^2\alpha}{\zeta^2}, 
\label{eq:B13}
\end{equation}
we obtain after some algebra
\begin{equation}
\tilde{\gamma}^2  = s\frac{[s^2\Delta^2 + 4s(1 + s)]^{1/2} - s\Delta(1 + 2s)}{2\zeta(1 + s)} .
\label{eq:B14}
\end{equation}
To calculate the maximum of function $\tilde{\gamma}^2(s)$ we have to find where its derivative is equal to zero. After some algebra the equation $d\tilde{\gamma}^2/ds = 0$ can be written as
\begin{eqnarray}
2\Delta^4(2s^6 \!&+&\! 7s^5 + 8s^4 + 3s^3) + 
   \Delta^2(16s^6 + 72s^5 + 119s^4 + 84s^3\nonumber\\
&+&\! 19s^2 - 2s) - (4s^4 + 20s^3 + 37s^2 + 30s + 9) = 0.
\label{eq:B15}
\end{eqnarray}
We consider this equation as a quadratic equation for $\Delta^2$\/. The roots of this equation have different signs. Then, taking into account that $\Delta^2 > 0$, we obtain that $s$ is defined by the equation
\begin{equation}
\Upsilon(s)  = \Delta^2,
\label{eq:B16}
\end{equation}
where $\Upsilon(s)$ is given by
\begin{eqnarray}
\Upsilon(s) &=& [s^2(2s^3 + 7s^2 + 8s + 3)]^{-1}\big[(256s^{10} + 2304s^9  
   \nonumber\\
&+& 9056s^8 + 20368s^7 + 28833s^6 + 26592s^5 \nonumber\\ 
&+& 15962s^4 + 6028s^3 + 1321s^2 + 140s + 4)^{1/2} \nonumber\\
&-& (16s^5 + 72s^4 + 119s^3 + 84s^2 + 19s - 2)\big] .
\label{eq:B17}
\end{eqnarray}
It is straightforward to obtain that $\Upsilon(s) \to \infty$ as $s \to 0$ and
$\Upsilon(s) \to 0$ as $s \to \infty$\/. We verified numerically that
$\Upsilon(s)$ is a monotonically decreasing function. Hence, Eq.~(\ref{eq:B16})
has the single solution $s_c$ for any value of $\Delta$\/. The dependence of
$s_c$ on $\Delta$ is shown in Fig.~\ref{fig:s_Delta}. 

Since $\tilde{\gamma}^2(0) = 0$ and $\tilde{\gamma}^2(s) \to -\infty$ as $s \to
\infty$ and it has only one extremum at $s = s_c$\/, this extremum is the
maximum, i.e.\ $\tilde{\gamma}_m = \tilde{\gamma}(s_c)$. The dependence of
$\Gamma = \zeta^{1/2}\tilde{\gamma}_n$ on $\Delta$ is shown in
Fig.~\ref{fig:gamma_m}.

Summarising the analysis we see that the function
$\tilde{\gamma}(\varphi,\kappa)$ has two local maxima. The first one is given by
Eq.~(\ref{eq:B7}) and it is taken at $\varphi = \pi/2$ and $\kappa$ given
by Eq.~(\ref{eq:B8}). The second local maximum is given by
Eq.~(\ref{eq:B14}) with $s = s_c$\/, and it is taken at $\varphi = \pi/2 +
\alpha$ and $\kappa = \zeta^{-1}s_c$\/, where $s_c$ is defined by
Eq.~(\ref{eq:B16}). We temporarily denote the first local maximum as
$\tilde{\gamma}_{m1}$ and the second as $\tilde{\gamma}_{m2}$\/. The absolute
maximum of $\tilde{\gamma}(\varphi,\kappa)$ is equal to the larger of the two
quantities $\tilde{\gamma}_{m1}$ and $\tilde{\gamma}_{m2}$\/. 

Equation~(\ref{eq:B7}) can be rewritten as 
\begin{equation}
\tilde{\gamma}_{m1} \approx \frac{\chi^2}{\zeta\Delta}\sqrt[4]{\frac4{27}}.
\label{eq:B18}
\end{equation}
Then it follows that 
\begin{equation}
\frac{\tilde{\gamma}_{m1}}{\tilde{\gamma}_{m2}} \approx 
   \frac{\chi^2}{\Delta\Gamma(\Delta)\sqrt\zeta}\sqrt[4]{\frac4{27}}
   \approx \frac{0.62\chi^2}{\Delta\Gamma(\Delta)\sqrt\zeta}.
\label{eq:B19}
\end{equation}
It is not difficult to obtain the approximate expressions
\begin{equation}
s_c \approx \left\{\begin{array}{cl}\displaystyle 
\frac1{2\Delta}, & \Delta \ll 1, \vspace*{3mm} \\
\displaystyle \frac1{\Delta\sqrt6}, & \Delta \gg 1,
\end{array}\right.
\label{eq:B20}
\end{equation}
Using this result we obtain
\begin{equation}
\Gamma \approx \left\{\begin{array}{cl}\displaystyle 
\frac1{2\sqrt{\Delta}}, & \Delta \ll 1, \vspace*{3mm} \\
\displaystyle \frac1\Delta\sqrt{\frac5{6\sqrt6}}, & \Delta \gg 1,
\end{array}\right.
\label{eq:B21}
\end{equation}
We verified numerically that $\Delta\Gamma(\Delta)$ is a monotonically increasing function of $\Delta$\/. Hence, it varies from $0.5\sqrt\Delta$ to 
$(5/6\sqrt6)^{1/2}$ when $\Delta$ varies from very small value to $\infty$\/.
Then it follows from Eq.~(\ref{eq:B19}) that $\tilde{\gamma}_{m1}/\tilde{\gamma}_{m2} \lesssim \chi^2(\zeta\Delta)^{-1/2}$\/. Hence, we conclude that $\tilde{\gamma}_{m1} \lesssim \tilde{\gamma}_{m2}$ for 
$\Delta > \chi^4/\zeta$\/. Since we assume that $\chi \simeq 1$, while the typical value of $\zeta$ is 100, we conclude that, for not very small values of $\Delta$ (say, $\Delta \gtrsim 0.1$), the absolute maximum of $\tilde{\gamma}$ is equal to $\tilde{\gamma}_{m2}$\/, i.e.\ $\tilde{\gamma}_m = \zeta^{-1/2}\Gamma(\Delta)$. It is taken at $\varphi \approx \pi/2 + \alpha$ and
$\kappa \approx \zeta^{-1}s_c(\Delta)$.


\begin{thebibliography}{30}
\expandafter\ifx\csname natexlab\endcsname\relax\def\natexlab#1{#1}\fi

\bibitem[{{Bucciantini} {et~al.}(2004){Bucciantini}, {Amato}, {Bandiera},
  {Blondin}, \& {Del Zanna}}]{bucc2004}
{Bucciantini}, N., {Amato}, E., {Bandiera}, R., {Blondin}, J.~M., \& {Del
  Zanna}, L. 2004, \aap, 423, 253

\bibitem[{{Chandrasekhar}(1961)}]{Chandrasekhar1961}
{Chandrasekhar}, S. 1961, {Hydrodynamic and hydromagnetic stability} (Oxford:
  Clarendon Press)

\bibitem[{{DeVore}(2012)}]{devore2012}
{DeVore}, C.~R. 2012, in American Astronomical Society Meeting Abstracts, Vol.
  220, American Astronomical Society Meeting Abstracts 220, 201.06

\bibitem[{{DeVore}(2013)}]{devore2013}
{DeVore}, C.~R. 2013, in AAS/Solar Physics Division Meeting, Vol.~44, AAS/Solar
  Physics Division Meeting, ...41
  
\bibitem[{{D\'{\i}az} {et~al.}(2012){D\'{\i}az}, {Soler}, \& {Ballester}}]{diaz2012}
{D\'{\i}az}, A. J., {Soler}, R., \& {Ballester}, J. L. 2012, \apj, 754, 41
   
\bibitem[{{Edwin} \& {Roberts}(1982)}]{Edwin1982}
{Edwin}, P.~M. \& {Roberts}, B. 1982, Solar Physics, 76, 239

\bibitem[{{Edwin} \& {Roberts}(1983)}]{Edwin1983}
{Edwin}, P.~M. \& {Roberts}, B. 1983, Solar Physics, 88, 179  

\bibitem[{{Goedbloed}(1971{\natexlab{a}})}]{goedbloed1971a}
{Goedbloed}, J.~P. 1971{\natexlab{a}}, Physica, 53, 412

\bibitem[{{Goedbloed}(1971{\natexlab{b}})}]{goedbloed1971b}
{Goedbloed}, J.~P. 1971{\natexlab{b}}, Physica, 53, 501

\bibitem[{{Goedbloed}(1971{\natexlab{c}})}]{goedbloed1971c}
{Goedbloed}, J.~P. 1971{\natexlab{c}}, Physica, 53, 535

\bibitem[{{Goedbloed} \& {Poedts}(2004)}]{GOEPOE2004}
{Goedbloed}, J.~P.~H. \& {Poedts}, S. 2004, {Principles of
  Magnetohydrodynamics} (Cambridge, UK: Cambridge University Press)

\bibitem[{{Hillier} {et~al.}(2012{\natexlab{a}}){Hillier}, {Berger}, {Isobe},
  \& {Shibata}}]{hillier2012a}
{Hillier}, A., {Berger}, T., {Isobe}, H., \& {Shibata}, K. 2012{\natexlab{a}},
  \apj, 746, 120

\bibitem[{{Hillier} {et~al.}(2011){Hillier}, {Isobe}, {Shibata}, \&
  {Berger}}]{hillier2011}
{Hillier}, A., {Isobe}, H., {Shibata}, K., \& {Berger}, T. 2011, \apjl, 736, L1

\bibitem[{{Hillier} {et~al.}(2012{\natexlab{b}}){Hillier}, {Isobe}, {Shibata},
  \& {Berger}}]{hillier2012b}
{Hillier}, A., {Isobe}, H., {Shibata}, K., \& {Berger}, T. 2012{\natexlab{b}},
  \apj, 756, 110

\bibitem[{{Isobe} {et~al.}(2005){Isobe}, {Miyagoshi}, {Shibata}, \&
  {Yokoyama}}]{isobe2005}
{Isobe}, H., {Miyagoshi}, T., {Shibata}, K., \& {Yokoyama}, T. 2005, \nat, 434,
  478

\bibitem[{{Isobe} {et~al.}(2006){Isobe}, {Miyagoshi}, {Shibata}, \&
  {Yokoyama}}]{isobe2006}
{Isobe}, H., {Miyagoshi}, T., {Shibata}, K., \& {Yokoyama}, T. 2006, \pasj, 58,
  423

\bibitem[{{Jones} \& {De Young}(2005)}]{jones2005}
{Jones}, T.~W. \& {De Young}, D.~S. 2005, \apj, 624, 586

\bibitem[{{Jun} \& {Norman}(1996)}]{jun1996}
{Jun}, B.-I. \& {Norman}, M.~L. 1996, \apj, 465, 800

\bibitem[{{Jun} {et~al.}(1995){Jun}, {Norman}, \& {Stone}}]{jun1995}
{Jun}, B.-I., {Norman}, M.~L., \& {Stone}, J.~M. 1995, \apj, 453, 332

\bibitem[{{Okamoto} {et~al.}(2007){Okamoto}, {Tsuneta}, {Berger}, {Ichimoto},
  {Katsukawa}, {Lites}, {Nagata}, {Shibata}, {Shimizu}, {Shine}, {Suematsu},
  {Tarbell}, \& {Title}}]{okamoto07}
{Okamoto}, T.~J., {Tsuneta}, S., {Berger}, T.~E., {et~al.} 2007, Science, 318,
  1577

\bibitem[{{O'Neill} {et~al.}(2009){O'Neill}, {De Young}, \&
  {Jones}}]{oneil2009}
{O'Neill}, S.~M., {De Young}, D.~S., \& {Jones}, T.~W. 2009, \apj, 694, 1317

\bibitem[{{Parker}(1974)}]{Parker1974}
{Parker}, E.~N. 1974, Solar Physics, 37, 127

\bibitem[{{Rayleigh}(1883)}]{Rayleigh1883}
{Rayleigh}, L. 1883, Proc. London Math. Soc., 14, 170

\bibitem[{{Roberts}(1981{\natexlab{a}})}]{Roberts1981a}
{Roberts}, B. 1981{\natexlab{a}}, Solar Physics, 69, 27

\bibitem[{{Robinson} {et~al.}(2004){Robinson}, {Dursi}, {Ricker}, {Rosner},
  {Calder}, {Zingale}, {Truran}, {Linde}, {Caceres}, {Fryxell}, {Olson},
  {Riley}, {Siegel}, \& {Vladimirova}}]{robinson2004}
{Robinson}, K., {Dursi}, L.~J., {Ricker}, P.~M., {et~al.} 2004, \apj, 601, 621

\bibitem[{{Ryutova} {et~al.}(2010){Ryutova}, {Berger}, {Frank}, {Tarbell}, \&
  {Title}}]{ryuetal10}
{Ryutova}, M., {Berger}, T., {Frank}, Z., {Tarbell}, T., \& {Title}, A. 2010,
  \solphys, 267, 75

\bibitem[{{Stone} \& {Gardiner}(2007)}]{stonegardiner2007}
{Stone}, J.~M. \& {Gardiner}, T. 2007, \apj, 671, 1726

\bibitem[{{Taylor}(1950)}]{Taylor1950}
{Taylor}, G.~I. 1950, Proc. Roy. Soc. London A, 201, 192

\bibitem[{{Terradas} {et~al.}(2008){Terradas}, {Arregui}, {Oliver}, \&
  {Ballester}}]{TERetal2008}
{Terradas}, J., {Arregui}, I., {Oliver}, R., \& {Ballester}, J.~L. 2008, \apjl,
  678, L153

\bibitem[{{Terradas} {et~al.}(2012){Terradas}, {Oliver}, \&
  {Ballester}}]{terretal2012}
{Terradas}, J., {Oliver}, R., \& {Ballester}, J.~L. 2012, \aap, 541, A102

\bibitem[{{Yang} {et~al.}(2011){Yang}, {Wang}, {Ye}, \& {Xue}}]{yang2011}
{Yang}, B.~L., {Wang}, L.~F., {Ye}, W.~H., \& {Xue}, C. 2011, Physics of
  Plasmas, 18, 072111

\bibitem[{{Zhang} {et~al.}(2012){Zhang}, {Lau}, {Rittersdorf}, {Weis},
  {Gilgenbach}, {Chalenski}, \& {Slutz}}]{zhangetal2012}
{Zhang}, P., {Lau}, Y.~Y., {Rittersdorf}, I.~M., {et~al.} 2012, Physics of
  Plasmas, 19, 022703

\end{thebibliography}
\end{document}